\documentclass{PoS}
\usepackage{pdflscape}
\usepackage{afterpage}
\usepackage{cite}
\usepackage{microtype}
\usepackage{amsmath}
\usepackage{mathtools}
\usepackage{stmaryrd}
\usepackage{accents}
\usepackage{tabulary}
\usepackage{booktabs}
\usepackage{multicol}
\usepackage{multirow}
\usepackage{float}
\usepackage{caption}
\usepackage{tikz}
\usetikzlibrary[matrix,patterns,calc]
\newcommand{\ubar}[1]{\underaccent{\bar}{#1}}

\title{\textbf{Exotic Branes in M-Theory}}
\author{David S. Berman$^a$ and Edvard T. Musaev$^{bc}$ \speaker{R. Otsuki}$^a$\\
\llap{$^a$}Queen Mary University of London,\\
Mile End Road, London, E1 4NS\\
\llap{$^b$}Moscow Institute of Physic and Technology,\\
Institutskii per. 9, Dolgoprudny, 141700, Russia\\
\llap{$^c$}Kazan Federal University,\\
Institute of Physics, Kremlevskaya 16a, Kazan, 420111, Kazan, Russia\\
E-mail: \email{d.s.berman@qmul.ac.uk}, \email{musaev.et@phystech.edu}, \email{r.otsuki@qmul.ac.uk}}
\abstract{\textbf{Abstract:}\hfill\\
We revisit curious objects in string and M-theory called \emph{exotic brane}---objects that are highly non-perturbative, possessing a tension that scales less than $g_s^{-2}$ and are generically of low-codimension. They are non-geometric in the sense that they are only well-defined locally as supergravity solutions and require duality transformations to patch correctly, in addition to the usual diffeomorphisms and gauge transformations.\par
We argue that Double Field Theory (DFT) and Exceptional Field Theory (EFT) are the prime setting in which to examine such objects. To emphasise this, we construct an explicit solution in $E_{7(7)} \times \mathbb{R}^+$ EFT that unifies many of the codimension-2 exotic branes into a single well-behaved solution on an extended spacetime. We further argue that there are in fact an infinite number of exotic branes in string- and M-theory, many of which fall into a more general class of exotic branes that do not afford even a local description in terms of conventional supergravity.
}
\FullConference{Corfu Summer Institute 2018 "School and Workshops on Elementary Particle Physics and Gravity"\\
		(CORFU2018)\\
		31 August - 28 September, 2018\\
		Corfu, Greece
}
\ShortTitle{Exotic Branes in Exceptional Field Theory}
\date{\today}
\begin{document}
\section{Introduction}
In recent years, there has been mounting evidence for the existence of branes beyond the conventional brane scan. It was argued in \cite{Obers:1998fb,deBoer:2012ma,deBoer:2010ud} that string theory also contains so-called `exotic branes' whose tensions\footnote{Note that we shall always work in the string frame unless explicitly stated otherwise.} scale with the string coupling as $g_s^{\alpha}$ with $\alpha <- 2$ (compare to D-branes or NS5-branes whose tensions scale as $\alpha = -1, -2$ respectively). Recent work includes the construction of worldvolume actions for non-geometric five-branes \cite{Kimura:2014upa,Kimura:2016anf,Chatzistavrakidis:2013jqa, Blair:2017hhy}, as well as instanton corrections to those exotic backgrounds. For a recent review of non-geometric backgrounds, see \cite{Plauschinn:2018wbo}\par
In addition to being highly non-perturbative, these branes are also generically of low codimension (2 or lower) and are consequently poorly understood objects which modify the asymptotic behaviour of the spacetime around it. The low codimension of these branes introduces an extra level of complexity. Since the transverse space is so small (or even non-existent) there is, in some sense, too little room to traverse freely around the brane. More precisely, such low-codimension branes possess a non-trivial $G(\mathbb{Z})$-monodromy\footnote{Here, $G(\mathbb{Z})$ denotes either a T-duality or U-duality group, depending on whether we are talking about a brane in Type II supergravity or M-theory} such that traversing around them only returns a state that is related to the original configuration by a $G(\mathbb{Z})$-transformation. Put another way, these are objects that require duality transformations to patch together, in addition to the usual diffeomorphisms and gauge transformations. As a consequence, these are not globally well-defined solutions of supergravity and even the definition of charge becomes troublesome already at codimension-2, as is discussed at length in \cite{deBoer:2012ma}. Such exotic branes thus include concrete realisations of the T-folds and U-folds originally proposed by Hull \cite{Hull:2006va}.\par
Due to the intimate link between exotic branes and dualities it is not unreasonable to expect the natural description of these exotic states to be in a theory in which the duality has been made manifest. There have been a number of related attempts at just such a formalism. The doubled worldsheet model \cite{Hull:2004in,Hull:2006va} was among the first of these that realised the duality explicitly by doubling the coordinates of a torus fibration to a $T^{2d}$ fibration upon which the duality acted naturally. Other attempts to construct a theory that makes T-duality manifest on the entire spacetime and not just on a fibration include `generalised geometries' \cite{Pacheco:2008ps,Coimbra:2011nw,Coimbra:2011ky,Coimbra:2012af} which are theories based on the observation that the parameters of the local symmetries of the bosonic sectors of 10- or 11-dimensional supergravity can be considered as sections of an extended tangent bundle which carries a natural $G(\mathbb{R})$ structure.\par
\emph{Extended Field Theories} (ExFT) take this one step further and realise these continuous symmetries as an explicit symmetry of a spacetime that has been extended by coordinates dual to the winding or wrappings modes of the branes in string- and M-theory. When $G (\mathbb{R}) = \operatorname{O}(n,n;\mathbb{R})$, the theory goes by the name of \emph{Double Field Theory} (DFT)\footnote{See \cite{Aldazabal:2013sca,Geissbuhler:2013uka,Berman:2013eva} for reviews} whereas the cases where $G (\mathbb{R}) = E_{n(n)}$ are called \emph{Exceptional Field Theories} (EFT). These have been constructed for the finite cases $n = 8, \ldots, 2$ in \cite{Hohm:2014fxa, Hohm:2013uia, Hohm:2013vpa, Abzalov:2015ega, Musaev:2015ces, Hohm:2015xna, Berman:2015rcc}, with further progress on the Kac-Moody algebra $E_{9(9)}$ made in \cite{Bossard:2017aae,Bossard:2018utw}\footnote{For the related $E_{11}$ scheme see, for example, \cite{West:2001as,West:2003fc,Tumanov:2015yjd,Tumanov:2016abm,Kleinschmidt:2003jf,Cook:2008bi,West:2004kb,West:2004iz,Cook:2009ri,Cook:2011ir}.}.\par
The groups $G(\mathbb{R})$, henceforth referred to as the \emph{solution-generation group}, that we realise as a manifest symmetry are evidently not the duality groups themselves but they do contain the discrete forms as a subgroup. More concretely, the two notions are related in the following manner. Since we have extended our spacetime with extra coordinates, we need some way to pick out the coordinates that we identify as the physical coordinates of string- or M-theory. In the presence of isometries, there is a ambiguity in the way we may pick out this `section' and the different spacetimes that we may pick out are related by genuine duality transformations. At the cost of realising $G(\mathbb{R})$ as an explicit symmetry, rather than the discrete duality groups, we obtain a theory that has proven to be remarkably versatile. In addition to providing a higher dimensional unification of Type II and 11-dimensional supergravity, one can obtain gauged supergravities through generalised Scherk-Schwarz reductions, or describe non-Riemannian backgrounds such as non-relativistic or ultra-relativistic limits  \cite{Cho:2018alk,Morand:2017fnv,Berman:2019izh}. However, the property that we shall focus on here is the apparent natural description of exotic backgrounds that ExFTs offer.\par
One of the strengths of ExFT is that it can naturally unify many branes into a single object on the extended space, just as the Type IIA--M-theory correspondence lead to the realisation that the D4 and NS5-branes descended from the same object in M-theory. The difference is that we may construct solutions that are entirely well-behaved on the extended spacetime, yet give rise to unusual (or even pathological) backgrounds upon reducing back down to 10- or 11-dimensions. Moreover, since the extended spaces are typically much larger than a single circle, we are afforded the unification of many more solutions of supergravity into a single solution in ExFT. The reasoning behind this is that, we may rotate the solution by $G$-transformations\footnote{In practise, we begin with a lift of a solution that is already on the charge lattice of BPS states and then restrict to transformations in $G(\mathbb{Z})$ to remain consistent with Dirac quantisation.} before reducing back to string- or M-theory. Provided that the internal space has a sufficient number of isometries, these new backgrounds will be related to the original by duality transformations. Examples of this may be found in \cite{Berkeley:2014nza,Blair:2016xnn,Berman:2014jsa,Bakhmatov:2016kfn,Kimura:2018hph,Blair:2014zba,Berman:2014hna,Bakhmatov:2017les,Berman:2019biz}. As powerful a technique this is, the real utility of this method lies in the fact that the parent solutions are indifferent to whether the solutions contained within it are geometric or non-geometric, as demonstrated in the DFT monopole.\par
The structure of this contribution is as follows. We shall begin with a brief introduction to the ideas behind EFTs in Section~\ref{sec:ExFT}, and then specialising to $E_{7(7)} \times \mathbb{R}^+$ in the latter parts. In Section~\ref{sec:Soln}, we describe a novel solution in $E_{7(7)} \times \mathbb{R}^+$ EFT that contains exotic branes. This is a more extreme example of how ExFTs treat the geometric and non-geometric objects on an equal footing; whilst the DFT monopole contains 2 geometric and 3 non-geometric solutions within it, the geometric sector of this solution consists only of the KK5A/B and KK6M. The remaining 16 branes are all of the non-geometric codimension-2 objects described in \cite{deBoer:2012ma}. Following this, in Section~\ref{sec:MoreBranes}, we answer the question of whether there are more exotic states that one might be able to describe with ExFT before closing with a discussion of the interpretation of these states in Section~\ref{sec:Discussion}.
\section{An $E_{n(n)} \times \mathbb{R}^+$ Exceptional Field Theory Primer}\label{sec:ExFT}
\subsection{The Symmetries of EFTs}
We now briefly cover some of the ideas underlying EFTs in general before specialising to $E_{7(7)}\times \mathbb{R}^+$ EFT. Just as general relativity can be understood as describing a $\operatorname{GL}(d)/\operatorname{SO}(d)$ coset, we describe the geometries appropriate for 11-dimensional supergravity by exceptional cosets of the form $E_{n(n)} / K(E_{n(n)})$, where $K(E_{n(n)})$ is the maximal compact subgroup of $E_{n(n)}$. EFTs realise a global $E_{n(n)}(\mathbb{R})$ symmetry on a spacetime that has been enlarged to accommodate representations of the exceptional group.\par
The starting point is to begin with a nominal splitting of an 11-dimensional space into a $d$-dimensional \emph{external space} $\mathcal{M}^d$ and the remainder $\mathcal{M}^n$ (such that $d+n =11$). The latter is extended to a space of dimension $\operatorname{dim}\rho_1$, where $\rho_1$ is a particular representation of $E_{n(n)}$ called the \emph{coordinate representation}. Schematically, we have $\mathcal{M}^{11} = \mathcal{M}^{d} \times \mathcal{M}^n \rightarrow \mathcal{M}^d \times \mathcal{M}^{\operatorname{dim} \rho_1}$. If we denote the coordinates of $\mathcal{M}^d$ as $x^\mu$ and the coordinates of $\mathcal{M}^{\operatorname{dim}\rho_1}$ as $Y^M$ (such that $\mu = 1, \ldots d$ and $M =1, \ldots, \operatorname{dim} \rho_1$), the extended coordinates $Y^M$ are acted on linearly by $G$. Of course, having added extra coordinates to the internal space, we must have some prescription of reducing back down to 11 or 10 dimensions. This is accomplished by the so-called \emph{section condition} which essentially restricts the coordinate dependence to a subset of the extended space. We shall postpone our discussion of it until after introducing the generalised Lie derivative.\par
In addition to the global symmetry, we also demand two local symmetries. The first are the residual diffeomorphisms on the external space $\mathcal{M}^d$ and the second is a \emph{generalised internal diffeomorphism}. Just as the infinitesimal diffeomorphisms of GR are generated by the Lie derivative we may, quite non-trivially, combine the local symmetries of supergravities in such a way as to describe them by a single \emph{generalised Lie derivative} $\mathbb{L}$ on the internal space whose parameters are given by a combination of the parameters of the internal diffeomorphisms with the gauge transformations of the form fields. We interpret this as a deformation of the usual Lie derivative\footnote{The present discussion only holds for $n \leq 7$; at $n=8$, the story is complicated by the introduction of a separate gauge symmetry such that the Lie derivative is parametrised by two objects $(\Lambda^M, \Sigma_M)$. For the present paper, we shall restrict our attention to $n \leq 7$ and we refer the reader to \cite{Berman:2019izh} for more details.} such that if the generalised internal coordinates transform as $\delta Y^M = - \Lambda^M$, then generalised vectors $V^M$ (of weight $\lambda(V)$) transform according to $\delta V^M  = {\mathbb{L}}_{\Lambda} V^M$. We \emph{define}
\begin{align}
{\mathbb{L}}_{\Lambda} V^M \coloneqq {[\Lambda, V]}^M + Y^{MN}{}_{KL} \partial_N \Lambda^K V^L + \left(\lambda (V)- \omega \right) \partial_N \Lambda^N V^M\,,
\end{align}
where $Y^{MN}{}_{KL}$ is the so-called \emph{Y-tensor} \cite{Berman:2012vc}, formed from $G$-invariants. In the final term, we see that the weight of the vector receives a correction by a \emph{universal weight} which, for EFTs, are given by $\omega = \frac{1}{n-2}$. The usual Lie derivative is recovered if one takes $G = \operatorname{GL}(d)$ and taking $Y^{MN}{}_{KL} = \omega = 0$. The fact that this identification is consistent hinges on re-interpreting the generalised Lie derivative as the sum of transport term and an adjoint-valued transformation,
\begin{align}
{\mathbb{L}}_{\Lambda} V^M \coloneqq \Lambda^N \partial_N V^M - \alpha_n  {\left({\mathbb{P}}_{\text{adj.}} \right)}^M{}_N{}^P{}_Q \partial_P \Lambda^Q V^N + \lambda (V) \partial_N \Lambda^N V^M\,,
\end{align}
where $\alpha_n$ are coefficients that depend on the particular theory and ${\mathbb{P}}_{\text{adj.}}$ is a projector from $\rho_1 \otimes {\bar{\rho}}_1$ onto the adjoint representation $\rho_{\text{adj.}}$. The story is quite general and is described in detail, alongside the case when $G= \operatorname{O}(n,n)$, in \cite{Aldazabal:2013mya,Coimbra:2011nw} amongst other places. The price that we pay for making this generalisation is that $\mathbb{L}$ no longer satisfies an algebra, nor is the action on a generalised vector antisymmetric. If one computes the commutator of two transformations, one finds an algebra obstructed by terms $\Delta_{\text{sec.}}$,
\begin{align}
[{\mathbb{L}}_U , {\mathbb{L}}_V ] = {\mathbb{L}}_{\llbracket U, V \rrbracket} + \Delta_{\text{sec.}}\,,
\end{align}
where $\llbracket U, V \rrbracket \coloneqq \frac{1}{2} ({\mathbb{L}}_U {\mathbb{L}}_V - {\mathbb{L}}_V {\mathbb{L}}_U)$ is called the \emph{E-bracket} (or \emph{C-bracket} in DFT). As it stands, the transformations do not close onto a well-defined algebra but this may be rectified if we demand the obstructions $\Delta_{\text{sec.}}$ vanishes by imposing some \emph{section condition}, by hand, to eliminate them. These terms arises from generic dependences of the fields and gauge parameters on all the coordinates $Y^M$ and it is sufficient to restrict this by demanding
\begin{align}\label{eq:Sect}
{(\partial \otimes \partial ) \vert}_{{\bar{\rho}}_2} = 0 \qquad \leftrightarrow \qquad Y^{MN}{}_{KL} \partial_M \otimes \partial_N = 0\,.
\end{align}
We call ${\bar{\rho}}_2 \subset {\bar{\rho}}_1 \otimes {\bar{\rho}}_1$ the \emph{section representation} of $G$. The notation above is understood to mean that the partial derivatives may either act on separate objects (\emph{strong section condition}) or on the same object (\emph{weak section condition}). There are only two inequivalent solutions to the section constraint and these generically pick out a $\operatorname{GL}(n)$ or $\operatorname{GL}(n-1) \times \operatorname{SL}(2)$ subgroup of $G$, corresponding to the \emph{M-theory section} or \emph{Type IIB section} respectively. Due to the large number of symmetries, it is possible to write down a unique 2-derivative action (or a pseudo-action in even dimensions, supplemented by a duality constraint) by demanding that all of these symmetries are respected and that it reduces correctly to both 11-dimensional supergravity and Type IIB supergravity upon solving the appropriate strong section condition. This will be given for $E_{7(7)}$ later, once we have covered the field content of these theories.
\subsection{The Field Content}
The bosonic field content of EFTs depend on the group under consideration but are all remarkably simple, consisting of only 
 \begin{align}
\{ g_{\mu \nu}, \mathcal{M}_{MN}, \mathcal{A}_\bullet{}^\bullet\}\,.
\end{align}
Each of these correspond to a metric on the external space (whose determinant we shall denote $g_{(d)}$), a \emph{generalised metric} on the extended internal space and a set of \emph{generalised Kaluza-Klein vectors} and an associated set of higher rank tensors (whose exact numbers and index structures we have represented with place-holders as they depend on the group $G$ under consideration). Here, just as the metric of GR parametrises the coset $\operatorname{GL}(d)/ \operatorname{SO}(d)$, the generalised metric parametrises the coset $(E_{n(n)} \times \mathbb{R}^+) / \operatorname{K}(E_{n(n)})$ and can be parametrised in terms of the supergravity fields, namely an internal $n$-dimensional metric and a set of form fields on the internal space $\mathcal{M}^n$. These have been constructed for all the finite cases in \cite{Berman:2011jh,Hillmann:2009ci,Godazgar:2013rja}. The extra scaling factor affords us the flexibility to take $\operatorname{det} \mathcal{M} \neq 1$ which we shall make use of later. For the solution presented here it is sufficient to note that, in the absence of internal potentials, the generalised metric takes on a diagonal form (see \eqref{eq:E7GenMetric}, for example).\par
The generalised Kaluza-Klein vectors act as the glue between the external and internal space, allowing for non-trivial fibrations of the latter over the former, and carry their own generalised gauge structure which must be compatible with the symmetries of the theory. This entails a much more intricate structure than one might first suppose and hinges on some rather delicate cancellations in the spirit of the tensor hierarchy of gauged supergravity. In particular, their generalised field strengths are not given by the na\"{i}ve generalised Lie-covariantised field strengths; they require a St\"{u}ckleberg-type correction, plus possibly additional auxiliary higher-form corrections, to ensure that is covariant with respect to the generalised diffeomorphisms. We shall not discuss the fine gauge structure in too much detail as it is not particularly relevant for the work presented here. However we note that, upon solving the section condition, the surviving components of these generalised gauge fields generically source the cross-sector components of the supergravity form-fields (and, additionally, the dual graviton in $E_{8(8)}$ EFT).\par
\subsection{Details of $E_{7(7)} \times \mathbb{R}^+$ EFT}
For our purposes, we are interested in the case $d=4$ such that we realise a global $E_{7(7)}\times \mathbb{R}^+$ symmetry\footnote{The $\mathbb{R}^+$ factor corresponds to the \emph{trombone symmetry} of supergravity---an extra off-shell scaling symmetry of supergravity Lagrangians, whose possible gaugings were worked out in \cite{LeDiffon:2008sh}} on an internal space of dimension $\operatorname{dim} \rho_1 = 56$. The coordinates on this internal extended space shall be denote $Y^M$ (where $M = 1, \ldots, 56$) that transform linearly under the fundamental representation $\rho_1 = \mathbf{56}$ of $E_{7(7)}$. The generalised Lie derivative is given by
\begin{align}
{\mathbb{L}}_\Lambda V^M & = {[\Lambda, V]}^M + Y^{MN}{}_{PQ} \partial_N \Lambda^P V^Q + \left( \lambda(V) - \frac{1}{2} \right) \partial_N \Lambda^N V^M\,,
\end{align}
where the $Y$-tensor for $E_{7(7)}$ is given in terms of its generators in the fundamental representation and an antisymmetric invariant $\Omega_{MN}$:
\begin{align}
Y^{MN}{}_{PQ} & = - 12 {(t_\alpha)}^{MN} {(t^\alpha)}_{PQ} - \frac{1}{2} \Omega^{MN} \Omega_{PQ}\,.
\end{align}
The index $\alpha = 1, \ldots, 133$ labels the adjoint representation. We may thus solve the section condition \eqref{eq:Sect} by demanding
\begin{align}
{(t^\alpha)}^{MN} \partial_M \otimes \partial_N = \Omega^{MN} \partial_M \otimes \partial_N = 0
\end{align}
The antisymmetric matrix $\Omega_{MN}$ is a weighted symplectic matrix \cite{Aldazabal:2013mya}, related to the symplectic matrix ${\tilde{\Omega}}_{MN}$ that it inherits from $\operatorname{Sp}(56) \supset E_{7(7)}$ by $\Omega_{MN} = e^{-\Delta} {\tilde{\Omega}}_{MN}$. Our conventions are to raise and lower fundamental indices with $\Omega^{MN}, \Omega_{MN}$ via
\begin{align}
V^M = \Omega^{MN} V_N , \qquad V_M = V^N \Omega_{NM} , \qquad \Omega_{MP} \Omega^{NP} = \delta^N_M\,,
\end{align}
from which it follows that ${(t_\alpha)}_{MN} \coloneqq {(t_\alpha)}_M{}^P \Omega_{PN} = {(t_\alpha)}_{NM}$. For other relations of the generators, we refer the reader to \cite{Hohm:2013uia}. This symplectic structure is compatible with the EFT structure such that
\begin{align}
\mathbb{L}_{\Lambda} \Omega_{MN} = 0, \qquad \mathcal{M}^{MN} = \Omega^{MP} \Omega^{NQ} \mathcal{M}_{PQ}
\end{align}
where $\mathcal{M}_{MN}$ is the generalised metric parametrising the coset $(E_{7(7)} \times \mathbb{R}^+)/\operatorname{SU}(8)$. The remaining fields of this theory are given by
\begin{align}
\{ g_{\mu \nu}, \mathcal{M}_{MN}, \mathcal{A}_{\mu}{}^M, {\mathcal{B}}_{\mu \nu, \alpha}, {\mathcal{B}}_{\mu \nu}{}^M \}\,,
\end{align}
where $g_{\mu \nu}$ is the external metric ($\mu, \nu = 1, \ldots, 4$) and $\mathcal{B}_{\mu \nu \bullet}$ are auxiliary gauge fields, required to covariantise the field strength of the generalised gauge fields $\mathcal{A}_\mu{}^M$. In the absence of internal potentials, and in the M-theory frame (i.e. decomposing under $\operatorname{GL}(7)$), the generalised metric, with internal potentials set to zero, is given solely in terms of the 7-dimensional internal metric $g_{mn}$ (determinant $g_{(7)}$) as
\begin{align}\label{eq:E7GenMetric}
{\mathcal{M}}_{MN} = e^{-\Delta} \operatorname{diag} [g_{(7)}^{\frac{1}{2}} g_{mn}; g_{(7)}^{\frac{1}{2}} g^{mn,pq}; g_{(7)}^{-\frac{1}{2}} g^{mn}; g_{(7)}^{-\frac{1}{2}} g_{mn,pq}]\,,
\end{align}
where $m, n = 1, \ldots, 7$ and $g_{mn,pq} \coloneqq g_{m[p} g_{q]n}$. The gauge field $\mathcal{A}_{\mu}{}^M$ has a na\"{i}ve field strength,
\begin{align}
F_{\mu \nu}{}^M & = 2 \partial_{[\mu} {\mathcal{A}}_{\nu]}{}^M - {\llbracket \mathcal{A}_{\mu}, \mathcal{A}_\nu \rrbracket}^M\,,
\end{align}
but this is not covariant under generalised diffeomorphisms. The full field strength is given by
\begin{align}
\mathcal{F}_{\mu \nu}{}^M = F_{\mu \nu}{}^M - 12 {(t^\alpha)}^{MN} \partial_N B_{\mu \nu \alpha} - \frac{1}{2} \Omega^{MK} B_{\mu \nu K}
\end{align}
which satisfies the Bianchi identity in terms of the 3-form field strengths of  the auxiliary fields
\begin{align}\label{eq:Bianchi}
3 {\mathcal{D}}_{[\mu} {\mathcal{F}}_{\nu \rho]}{}^M = - 12 {(t^\alpha)}^{MN} \partial_N {\mathcal{H}}_{\mu \nu \rho \alpha} - \frac{1}{2} \Omega^{MN} {\mathcal{H}}_{\mu \nu \rho N}\,.
\end{align}
Here, the covariant derivative is the \emph{Lie-covariantised derivative} $\mathcal{D}_\mu \coloneqq \partial_\mu - \mathbb{L}_{\mathcal{A}_\mu}$. Additionally, in $d=4$, there is an associated twisted self-duality relation of the field strength, given by
\begin{align}\label{eq:Duality}
{\mathcal{F}}_{\mu \nu}{}^M = - \frac{1}{2} \sqrt{-g_{(4)}} \varepsilon_{\mu \nu \rho \sigma}g^{\rho \lambda} g^{\sigma \tau} \Omega^{MN} {\mathcal{M}}_{NK} {\mathcal{F}}_{\lambda \tau}{}^K\,,
\end{align}
where $\varepsilon_{\mu \nu \rho \sigma}$ is the alternating symbol in four dimensions. This supplements the equations of motion obtained from the pseudo-action
\begin{align}
\begin{aligned}
S_{E_{7(7)}} = & \int \textrm{d}^4 x \textrm{d}^{56} Y \left( {\mathcal{L}}_{\text{E-H}} + {\mathcal{L}}_{\text{sc.}} + {\mathcal{L}}_{\text{Y-M}} - e_{(4)} V \right) + S_{\text{top.}}\,, \qquad e_{(4)} = \sqrt{-g_{(4)}}\,,\\
{\mathcal{L}}_{\text{E-H}} = & e_{(4)} \hat{R}\\
{\mathcal{L}}_{\text{sc.}} = & \frac{1}{48} e_{(4)} g^{\mu \nu} {\mathcal{D}}_\mu {\mathcal{M}}_{MN} {\mathcal{D}}_\nu {\mathcal{M}}^{MN}\\
{\mathcal{L}}_{\text{Y-M}} = & - \frac{1}{8} e_{(4)} {\mathcal{M}}_{MN} {\mathcal{F}}^{\mu \nu M} {\mathcal{F}}_{\mu \nu}{}^N\\
\phantom{V =} & 
\begin{aligned}
\mathllap{V =} & - \frac{1}{48} {\mathcal{M}}^{MN} \partial_M \mathcal{M}^{KL} \partial_N \mathcal{M}_{KL} + \frac{1}{2} \mathcal{M}^{MN} \partial_M \mathcal{M}^{KL} \partial_K \mathcal{M}_{NL}\\
& - \frac{1}{2}  \partial_M \ln |g_{(4)}| \partial_N \mathcal{M}^{MN} - \frac{1}{4} \mathcal{M}^{MN} \partial_M \ln |g_{(4)}| \partial_N \ln |g_{(4)}| - \frac{1}{4} \mathcal{M}^{MN}\partial_M g^{\mu \nu} \partial_N g_{\mu \nu}
\end{aligned}\\
S_{\text{top.}} = & - \frac{1}{24} \int \textrm{d}^5 x \int \textrm{d}^{56} Y \varepsilon^{\mu \nu \rho \sigma \tau} {\mathcal{F}}_{\mu \nu}{}^M {\mathcal{D}}_{\rho} {\mathcal{F}}_{\sigma \tau M}\,.
\end{aligned}
\end{align}
\section{A Novel Solution in $E_{7(7)}\times \mathbb{R}^+$ EFT}\label{sec:Soln}
We are now equipped to discuss the non-geometric solution presented in \cite{Berman:2018okd}. It adds to a growing list of solutions in DFT\cite{Berkeley:2014nza,Blair:2016xnn,Berman:2014jsa,Bakhmatov:2016kfn,Kimura:2018hph} and the various EFTs\cite{Blair:2014zba,Berman:2014hna,Bakhmatov:2017les} in which a single solution in the extended space reduces to a number of distinct solutions upon applying the section constraint. This phenomenon can be thought of as a direct analogue of the remarkable unification of, for example, the D4- and NS5A-branes in Type IIA string theory into the M5-brane of M-theory. The only difference is that we have a much larger choice of ways in which one can `reduce' the EFT solution (not just transverse or longitudinal to the brane as in the case of the M5-brane). It is hopefully self-evident that the number of branes a given EFT can unify is limited primarily by the size of the extended space and we shall comment on this later in Section~\ref{sec:Discussion}.\par
The solution presented here is a very close analogue of, what we shall call, the \emph{geometric solution} presented in \cite{Berkeley:2014nza} in which they also constructed a solution in $E_{7(7)}$ EFT (though one that covered the majority of the conventional branes) and we have summarised the set of branes that each of these solutions covers is Figure \ref{fig:Brane}.
\afterpage{
\thispagestyle{empty}
\begin{landscape}
\begin{figure}[hbtp]
\centering
\begin{tikzpicture}
\matrix(M)[ampersand replacement=\&, matrix of math nodes, row sep=3em, column sep=2 em, minimum width=2em]{
\clap{\phantom{a}}\&\&\&\&\&\&\&\&\&\clap{\phantom{a}}\&\clap{\phantom{a}}\&\&\&\clap{\phantom{a}}\\
\clap{\text{M:}}\&\clap{\phantom{a}}\&\clap{\text{PM}}\&\&\clap{\text{M2}}\&\&\&\&\&\clap{\text{M5}}\&\&\clap{\text{KK6}}\&\clap{\phantom{a}}\&\\
\clap{\text{IIA:}}\&\clap{\text{WA}}\&\clap{\text{F1A}}\&\clap{\text{D0}}\&\&\clap{\text{D2}}\&\&\clap{\text{D4}}\&\&\clap{\text{D6}}\&\&\clap{\text{NS5A}}\&\clap{\text{KK5A}}\&\\
\clap{\text{IIB:}}\&\clap{\text{WB}}\&\clap{\text{F1B}}\&\&\clap{\text{D1}}\&\&\clap{\text{D3}}\&\&\clap{\text{D5}}\&\clap{\phantom{a}}\&\clap{\text{D7}}\&\clap{\text{NS5B}}\&\clap{\phantom{a}}\&\clap{\phantom{a}}\\
\clap{\phantom{a}}\&\&\&\&\&\&\&\&\&\clap{\phantom{a}}\&\clap{\phantom{a}}\&\clap{\phantom{a}}\&\clap{\text{KK5B}}\&\clap{\phantom{a}}\\
\clap{\text{IIB:}}\&\mathclap{0_4^{(1,6)}\text{B}}\&\mathclap{1_4^6\text{B}}\&\&\mathclap{1_3^6}\&\&\mathclap{3_3^4}\&\&\mathclap{5_3^2}\&\mathclap{\phantom{a}}\&\mathclap{7_3}\&\mathclap{5_2^2\text{B}}\&\mathclap{\phantom{a}}\&\mathclap{\phantom{a}}\\
\clap{\text{IIA:}}\&\mathclap{0_4^{(1,6)}\text{A}}\&\mathclap{1_4^6\text{A}}\&\mathclap{0_3^7}\&\&\mathclap{2_3^5}\&\&\mathclap{4_3^3}\&\&\mathclap{6_3^1}\&\mathclap{\phantom{a}}\&\mathclap{5_2^2\text{A}}\&\clap{\text{KK5A}}\&\\
\clap{\text{M:}}\&\clap{\phantom{a}}\&\mathclap{0^{(1,7)}}\&\&\mathclap{2^6}\&\&\&\&\&\mathclap{5^3}\&\&\clap{\text{KK6}}\&\clap{\phantom{\text{a}}}\&\clap{\phantom{a}}\\
\clap{\phantom{a}}\&\&\&\&\&\&\&\&\&\clap{\phantom{a}}\&\clap{\phantom{a}}\&\&\clap{\phantom{a}}\&\\
	};
\draw[latex-latex, draw=blue, fill=blue] (M-2-3) -- (M-3-2);
\draw[latex-latex, draw=blue, fill=blue] (M-2-3) -- (M-3-4);
\draw[latex-latex, draw=blue, fill=blue] (M-2-5) -- (M-3-3);
\draw[latex-latex, draw=blue, fill=blue] (M-2-5) -- (M-3-6);
\draw[latex-latex, draw=blue, fill=blue] (M-2-10) -- (M-3-8);
\draw[latex-latex, draw=blue, fill=blue] (M-2-10) -- (M-3-12);
\draw[latex-latex, draw=blue, fill=blue] (M-2-12) -- (M-3-10);
\draw[latex-latex, draw=blue, fill=blue] (M-2-12) -- (M-3-13);
\draw[latex-latex, draw=red, fill=red] (M-3-2) -- (M-4-2);
\draw[latex-latex, draw=red, fill=red] (M-3-3) -- (M-4-3);
\draw[latex-latex, draw=red, fill=red] (M-3-2) -- (M-4-3);
\draw[latex-latex, draw=red, fill=red] (M-3-3) -- (M-4-2);
\draw[latex-latex, draw=red, fill=red] (M-3-4) -- (M-4-5);
\draw[latex-latex, draw=red, fill=red] (M-4-5) -- (M-3-6);
\draw[latex-latex, draw=red, fill=red] (M-3-6) -- (M-4-7);
\draw[latex-latex, draw=red, fill=red] (M-4-7) -- (M-3-8);
\draw[latex-latex, draw=red, fill=red] (M-3-8) -- (M-4-9);
\draw[latex-latex, draw=red, fill=red] (M-4-9) -- (M-3-10);
\draw[latex-latex, draw=red, fill=red] (M-3-10) -- (M-4-11);
\draw[latex-latex, draw=red, fill=red] (M-3-12) -- (M-4-12);
\draw[latex-latex, draw=red, fill=red] (M-3-13) -- (M-5-13);
\draw[latex-latex, draw=red, fill=red] (M-3-12) -- (M-5-13);
\draw[latex-latex, draw=red, fill=red] (M-3-13) -- (M-4-12);
\draw[latex-latex] (M-4-3) to[bend right,min distance=5mm] (M-4-5);
\draw[latex-latex] (M-4-9) to[bend right,min distance=5mm] (M-4-12);
\draw[latex-latex] (M-4-7) to[out=-135,in=-45,min distance=10mm](M-4-7);
\draw[latex-latex] (M-4-11) to[bend left] (M-6-11);
\draw[latex-latex] (M-6-7) to[out=45,in=135, min distance=10mm](M-6-7);
\draw[latex-latex] (M-6-9) to[bend left,min distance=5mm] (M-6-12);
\draw[latex-latex] (M-6-3) to[bend left,min distance=5mm] (M-6-5);
\draw[latex-latex, draw=red, fill=red] (M-6-2) -- (M-7-2);
\draw[latex-latex, draw=red, fill=red] (M-6-3) -- (M-7-3);
\draw[latex-latex, draw=red, fill=red] (M-6-2) -- (M-7-3);
\draw[latex-latex, draw=red, fill=red] (M-6-3) -- (M-7-2);
\draw[latex-latex, draw=red, fill=red] (M-7-4) -- (M-6-5);
\draw[latex-latex, draw=red, fill=red] (M-6-5) -- (M-7-6);
\draw[latex-latex, draw=red, fill=red] (M-7-6) -- (M-6-7);
\draw[latex-latex, draw=red, fill=red] (M-6-7) -- (M-7-8);
\draw[latex-latex, draw=red, fill=red] (M-7-8) -- (M-6-9);
\draw[latex-latex, draw=red, fill=red] (M-6-9) -- (M-7-10);
\draw[latex-latex, draw=red, fill=red] (M-7-10) -- (M-6-11);
\draw[latex-latex, draw=red, fill=red] (M-6-12) -- (M-7-12);
\draw[latex-latex, draw=red, fill=red] (M-5-13) -- (M-7-12);
\draw[latex-latex, draw=red, fill=red] (M-5-13) -- (M-7-13);
\draw[latex-latex, draw=red, fill=red] (M-6-12) -- (M-7-13);
\draw[latex-latex, draw=blue, fill=blue] (M-8-3) -- (M-7-2);
\draw[latex-latex, draw=blue, fill=blue] (M-8-3) -- (M-7-4);
\draw[latex-latex, draw=blue, fill=blue] (M-8-5) -- (M-7-3);
\draw[latex-latex, draw=blue, fill=blue] (M-8-5) -- (M-7-6);
\draw[latex-latex, draw=blue, fill=blue] (M-8-10) -- (M-7-8);
\draw[latex-latex, draw=blue, fill=blue] (M-8-10) -- (M-7-12);
\draw[latex-latex, draw=blue, fill=blue] (M-8-12) -- (M-7-10);
\draw[latex-latex, draw=blue, fill=blue] (M-8-12) -- (M-7-13);
\pattern[pattern=north east lines, pattern color=orange, draw opacity=0.6]
	($(M-1-1.south east)!0.5!(M-2-2.north west)$)
	--($(M-5-1.north east)!0.5!(M-4-2.south west)$)
	--($(M-5-10.north east)!0.5!(M-4-11.south west)$)
	--($(M-3-10.south west)!0.5!(M-4-11.north east)$)
	--($(M-4-11.north east)!0.5!(M-3-12.south west)$)
	--($(M-5-11.north east)!0.5!(M-4-12.south west)$)
	--($(M-5-12.north east)!0.5!(M-4-13.south west)$)
	--($(M-5-13.south west)!0.5!(M-6-12.north east)$)
	--($(M-6-14.north west)!0.5!(M-5-13.south east)$)
	--($(M-2-13.north east)!0.5!(M-1-14.south west)$)
	--($(M-1-1.south east)!0.5!(M-2-2.north west)$);
\pattern[pattern=north west lines, pattern color=green, draw opacity=0.6]
	($(M-5-1.south west)!0.5!(M-6-2.north east)$)
	--($(M-9-1.south west)!0.5!(M-8-2.north east)$)
	--($(M-9-13.north east)!0.5!(M-8-14.south west)$)
	--($(M-4-13.south east)!0.5!(M-5-14.north west)$)
	--($(M-5-12.north east)!0.5!(M-4-13.south west)$)
	--($(M-6-12.north east)!0.5!(M-5-13.south west)$)
	--($(M-6-11.north east)!0.5!(M-5-12.south west)$)
	--($(M-7-12.north west)!0.5!(M-6-11.south east)$)
	--($(M-7-10.north east)!0.5!(M-6-11.south west)$)
	--($(M-5-10.south east)!0.5!(M-6-11.north west)$)
	--($(M-5-1.south west)!0.5!(M-6-2.north east)$);
\end{tikzpicture}
\caption{A comparison of the objects covered by the geometric solution \cite{Berkeley:2014nza} (backed by the orange hashing) and non-geometric solution (backed by the green hashing) out of all the branes down to codimension-2 and $0 \leq \alpha \leq -2$, as well as their M-theory lifts. Red lines denote T-duality, blue lines denote lifts/reductions and black lines denote S-dualities.}
\label{fig:Brane}
\end{figure}
\end{landscape}
}
However, in order to understand the figure, we need to clarify the notation. We adopt the same conventions that have become the standard for describing exotic branes in the literature in which we refer to branes by the structure of their mass formulae, given in terms of the radii of any internal directions that they wrap and the constants of string- or M-theory, as follows
\begin{align}
\text{Type II}: && \text{M}(b_n^{(\ldots, d,c)}) & = \frac{\ldots {(R_{k_1} \ldots R_{k_d})}^3 {(R_{j_1} \ldots R_{j_c})}^2 {(R_{i_1} \ldots R_{i_b})}}{g_s^n l_s^{1 + b + 2c + 3d + \ldots}},\\
\text{M-Theory}: && \text{M}(b^{(\ldots, d,c)}) & = \frac{\ldots {(R_{k_1} \ldots R_{k_d})}^3 {(R_{j_1} \ldots R_{j_c})}^2 {(R_{i_1} \ldots R_{i_b})}}{  l_p^{1 + b + 2c + 3d + \ldots}}.
\end{align}
In the notation presented here, the familiar branes take on the following designations
\begin{multicols}{2}
\raggedcolumns
\noindent
\begin{align}
\begin{array}{ccc}
\text{F1} & \longleftrightarrow & 1_0\\
\text{D}$p$\ & \longleftrightarrow & p_1\\
\text{NS5} & \longleftrightarrow & 5_2\\
\text{KK5} & \longleftrightarrow & 5_2^1\\
\end{array}
\end{align}
\columnbreak
\begin{align}
\begin{array}{ccc}
\text{WM} & \longleftrightarrow & 0\\
\text{M}2 & \longleftrightarrow & 2\\
\text{M}5 & \longleftrightarrow & 5\\
\text{KK}6 & \longleftrightarrow & 6^1\\
\end{array}
\end{align}
\end{multicols}
Of these standard branes only the KK-monopoles in 10 and 11 dimensions have a distinguished direction, namely the circle of the Hopf fibration, and it is this direction that appears quadratically in the mass formula. We shall also, on occasion, annotate the names of the branes with the theory for convenience e.g. the Kaluza-Klein monopole in Type IIA shall be denoted $5_2^1$A.\par
The non-geometric solution itself is remarkably simple in form and we outline the ansatz below. For the 4-dimensional external space, we adopt cylindrical coordinates plus time such that $x^\mu = (t, r, \theta, z)$ and the base external metric for both the M-theory and Type IIB section is taken to be 
\begin{align}\label{eq:Ext}
g_{\mu \nu} = \operatorname{diag} [ - {(HK^{-1})}^{-\frac{1}{2}}, {(HK)}^{\frac{1}{2}}, r^2 {(HK)}^{\frac{1}{2}}, {(HK^{-1})}^{\frac{1}{2}}] = {\hat{g}}_{\hat{\mu} \hat{\nu}}\,,
\end{align}
where $H$ denotes the harmonic function in the $r$-$\theta$ plane and $K$ is a function commonly defined for codimension-2 exotic states
\begin{align}
H(r) = h_0 + \sigma \ln \frac{\mu}{r}\,, \qquad K = H^2 + \sigma^2 \theta^2\,.
\end{align}
Here, $h_0$ and $\sigma$ are constants and $\mu$ will become the charge of the objects. In all the solutions that we obtain, $g_{\mu \nu}$ gets dressed and we shall denote the external metric of a particular solution with a superscript e.g. in the $5^3$ frame, we have $g_{\mu \nu}^{5^3} = {(HK^{-1})}^{\frac{1}{6}} g_{\mu\nu}$. The exact prescription shall be outlined shortly.\par
For the generalised metric, every solution that we obtain will not have any internal potentials such that it is diagonal in every frame. Its 56 components are given by 27 components of ${(HK^{-1})}^{\frac{1}{2}}$, 27 components of ${(H^{-1})}^{-\frac{1}{2}}$ and one component each of ${(HK^{-1})}^{\frac{3}{2}}$ and ${(HK^{-1})}^{-\frac{3}{2}}$. The latter two directions will distinguish two of the internal directions. Because we consider an extra $\mathbb{R}^+$ factor, we are afforded the flexibility of introducing an extra scale factor such that the combination of the scale factor and internal metric \emph{in that frame} is invariant in a manner that we shall clarify later.\par
Finally, for the gauge fields, we take the auxiliary fields ${\mathcal{B}}_{\mu \nu \bullet} = 0$ and take ${\mathcal{A}}_\mu{}^M$ to be
\begin{align}\label{eq:PotentialAnsatz}
{\mathcal{A}}_\mu{}^M & = ( - H^{-1} K a^M ,\, 0,\, 0,\,  - K^{-1} \theta \sigma {\tilde{a}}^M)\,.
\end{align}
The two generalised vectors $a^M$ and ${\tilde{a}}^M$ determine the direction in which the vector points in the internal space. Note that the duality condition \eqref{eq:Duality} means that they are not actually independent. It is straightforward to verify that its field strength (which reduces to the Abelian field strength under this ansatz) also satisfies the Bianchi identity \eqref{eq:Bianchi}.
\subsection{An Example: Rotating the $5^3$ to the $2^6$}
To find the M-theory branes we first decompose under $\operatorname{GL}(7)$. If the usual internal coordinates are given by
\begin{align}
y^m =  (\xi, \chi, w^a) = (Y^\xi, Y^\chi, Y^a)\,,\qquad a = 1, \ldots, 5\,,
\end{align}
the 56 extended coordinates are split into
\begin{align}\label{eq:IntCoords}
Y^M = ( Y^\xi, Y^\chi, Y^a; Y_{\xi \chi}, Y_{\xi a}, Y_{\chi a}, Y_{ab}; Y_{\xi} Y_{\chi}, Y_a; Y^{\xi a} Y^{\chi a}, Y^{ab})\,.
\end{align}
For the generalised metric, we seek some particular frame $\textrm{M}$ which admit an external metric $g_{\mu \nu}^{\textrm{M}}$ and a 7-dimensional internal metric $g_{mn}^{\textrm{M}}$ such that $\mathcal{M}_{MN}$ conforms with the structure \eqref{eq:E7GenMetric} according to
\begin{align}\label{eq:NonGeom}
\begin{aligned}
\mathcal{M}_{MN} & = {\left|g^{5^3}_{(4)}\right|}^{-\frac{1}{4}} \operatorname{diag} [ g^{5^3}_{mn}; g_{5^3}^{mn,pq}; {(g^{5^3}_{(7)})}^{-1} g_{5^3}^{mn}; {(g^{5^3}_{(7)})}^{-1} g^{5^3}_{mn,pq}]\\
	& = {\left|g^{2^6}_{(4)}\right|}^{-\frac{1}{4}} \operatorname{diag} [ g^{2^6}_{mn}; g_{2^6}^{mn,pq}; {(g^{2^6}_{(7)})}^{-1} g_{2^6}^{mn}; {(g^{2^6}_{(7)})}^{-1} g^{2^6}_{mn,pq}]\\
	& = {\left|g^{0^{(1,7)}}_{(4)}\right|}^{-\frac{1}{4}} \operatorname{diag} [ g^{0^{(1,7)}}_{mn}; g_{0^{(1,7)}}^{mn,pq}; {(g^{0^{(1,7)}}_{(7)})}^{-1} g_{0^{(1,7)}}^{mn}; {(g^{0^{(1,7)}}_{(7)})}^{-1} g^{0^{(1,7)}}_{mn,pq}]\\
	& = {\left|g^{\text{KK6}}_{(4)}\right|}^{-\frac{1}{4}} \operatorname{diag} [ g^{\text{KK6}}_{mn}; g_{\text{KK6}}^{mn,pq}; {(g^{\text{KK6}}_{(7)})}^{-1} g_{\text{KK6}}^{mn}; {(g^{\text{KK6}}_{(7)})}^{-1} g^{\text{KK6}}_{mn,pq}]\,,\\
	& \qquad  \qquad \qquad \qquad \vdots
\end{aligned}
\end{align}
where the ellipsis denotes analogous decompositions for the Type IIA/B branes that we obtain. Similarly, the generalised vector $\mathcal{A}_{\mu}{}^M$ splits into
\begin{align}
\mathcal{A}_\mu{}^M \rightarrow (\mathcal{A}_\mu{}^m, \mathcal{A}_{\mu, mn}, \mathcal{A}_{\mu m}, \mathcal{A}_{\mu}{}^{mn})\,.
\end{align}
When non-zero, the first and third of these will be re-interpreted as a (conventional) Kaluza-Klein vectors and dual graviton components whilst the remaining components will source the cross-sector M-theory potentials $\mathcal{A}_{\mu mn} \sim A_{\mu mn}$ and ${\mathcal{A}}_{\mu}{}^{mn} \sim \epsilon^{mn p_1 \ldots p_5} A_{\mu p_1 \ldots p_5}$.\par
As an example, we consider the rotation from the $5^3$ frame to the $2^6$ frame. We begin with the generalised metric
\begin{align}
\phantom{\mathcal{M}_{MN}} &
\begin{alignedat}{2}\label{eq:53GM}
\mathllap{\mathcal{M}_{MN}} = {|g_{(4)}|}^{-\frac{1}{4}} \operatorname{diag}[
& {(HK^{-1})}^{\frac{1}{2}}, {(HK^{-1})}^{\frac{1}{2}}, {(HK^{-1})}^{-\frac{1}{2}} \delta_{(5)};\\
& {(HK^{-1})}^{-\frac{3}{2}}, {(HK^{-1})}^{-\frac{1}{2}} \delta_{(5)}, {(HK^{-1})}^{-\frac{1}{2}} \delta_{(5)}, {(HK^{-1})}^{\frac{1}{2}} \delta_{(10)};\\
& {(HK^{-1})}^{-\frac{1}{2}}, {(HK^{-1})}^{-\frac{1}{2}}, {(HK^{-1})}^{\frac{1}{2}} \delta_{(5)};\\
& {(HK^{-1})}^{\frac{3}{2}}, {(HK^{-1})}^{\frac{1}{2}} \delta_{(5)}, {(HK^{-1})}^{\frac{1}{2}} \delta_{(5)}, {(HK^{-1})}^{-\frac{1}{2}} \delta_{(10)}]
\end{alignedat}\\
&
\begin{alignedat}{2}\label{eq:53GenMetric}
= \smash{\underbrace{ \left[{(HK^{-1})}^{-\frac{1}{6}} {| g_{(4)}|}^{-\frac{1}{4}} \right]}_{{\left|g_{(4)}^{5^3}\right|}^{-\frac{1}{4}}} \operatorname{diag} [}
& {(HK^{-1})}^{\frac{2}{3}}, {(HK^{-1})}^{\frac{2}{3}}, {(HK^{-1})}^{-\frac{1}{3}} \delta_{(5)};\\
& {(HK^{-1})}^{-\frac{4}{3}}, {(HK^{-1})}^{-\frac{1}{3}} \delta_{(5)}, {(HK^{-1})}^{-\frac{1}{3}} \delta_{(5)}, {(HK^{-1})}^{\frac{2}{3}} \delta_{(10)};\\
& {(HK^{-1})}^{-\frac{1}{3}}, {(HK^{-1})}^{-\frac{1}{3}}, {(HK^{-1})}^{\frac{2}{3}} \delta_{(5)};\\
& {(HK^{-1})}^{\frac{5}{3}}, {(HK^{-1})}^{\frac{2}{3}} \delta_{(5)}, {(HK^{-1})}^{\frac{2}{3}} \delta_{(5)}, {(HK^{-1})}^{-\frac{1}{3}} \delta_{(10)}]\,.
\end{alignedat}
\end{align}
The first line shows the components of the generalised metric as described above \eqref{eq:PotentialAnsatz} whilst the second line is the form specialised to the $5^3$ frame. One may verify that this is indeed consistent with the form \eqref{eq:E7GenMetric} if one identifies
\begin{align}
g_{mn}^{5^3} & = \operatorname{diag} [ {(HK^{-1})}^{\frac{2}{3}}, {(HK^{-1})}^{\frac{2}{3}}, {(HK^{-1})}^{-\frac{1}{3}} \delta_{(5)}]\,,\\
{\left| g_{(4)}^{5^3}\right|}^{-\frac{1}{4}} & \coloneqq \left[ {(HK^{-1})}^{-\frac{1}{6}} {| g_{(4)}|}^{-\frac{1}{4}} \right] \Rightarrow g_{\mu \nu}^{5^3} = {(HK^{-1})}^{\frac{1}{6}} g_{\mu \nu}\,.
\end{align}
In this frame, we choose the potentials to point outside of the section such that they source the cross-sector potentials of the $5^3$ according to
\begin{align}\label{eq:53EFTVec}
\mathcal{A}_t{}^{\xi \chi} & = - H^{-1}K\,,\\
\mathcal{A}_{z,\xi \chi} & = - K^{-1} \theta \sigma\,.
\end{align}
Upon taking section, we may combine the data to end up with the background of the $5^3$-brane
\begin{gather}\label{eq:53Metric}
\begin{gathered}
\textrm{d} s^2_{5^3} = {(HK^{-1})}^{-\frac{1}{3}} ( -\textrm{d} t^2 + \textrm{d} {\vec{w}}^2_{(5)} ) + {(HK^{-1})}^{\frac{2}{3}} ( \textrm{d} z^2 + \textrm{d} \xi^2 + \textrm{d} \chi^2) + H^{\frac{2}{3}} K^{\frac{1}{3}} ( \textrm{d} r^2 + r^2 \textrm{d} \theta^2)\,,\\
A_{(3)} = -K^{-1} \theta \sigma \textrm{d} z \wedge \textrm{d} \xi \wedge \textrm{d} \chi\,, \qquad A_{(6)} = - H^{-1}K \textrm{d} t \wedge \textrm{d} w^1 \ldots \wedge \textrm{d} w^5\,.\\
\end{gathered}
\end{gather}
If we apply the rotation $Y^m \leftrightarrow Y_m, Y_{mn} \leftrightarrow Y^{mn}$, and reshuffle various factors of ${(HK^{-1})}$, we obtain the generalised metric
\begin{align}
\begin{aligned}\label{eq:26GenMetric}
\mathcal{M}_{MN} = \left[ {(HK^{-1})}^{\frac{1}{6}}  {|g_{(4)}|}^{-\frac{1}{4}} \right] \operatorname{diag} [
& {(HK^{-1})}^{-\frac{2}{3}}, {(HK^{-1})}^{-\frac{2}{3}}, {(HK^{-1})}^{\frac{1}{3}} \delta_{(5)};\\
& {(HK^{-1})}^{\frac{4}{3}}, {(HK^{-1})}^{\frac{1}{3}} \delta_{(5)}, {(HK^{-1})}^{\frac{1}{3}} \delta_{(5)}, {(HK^{-1})}^{-\frac{2}{3}} \delta_{(10)};\\
& {(HK^{-1})}^{\frac{1}{3}}, {(HK^{-1})}^{\frac{1}{3}}, {(HK^{-1})}^{-\frac{2}{3}} \delta_{(5)};\\
& {(HK^{-1})}^{-\frac{5}{3}}, {(HK^{-1})}^{-\frac{2}{3}} \delta_{(5)}, {(HK^{-1})}^{-\frac{2}{3}} \delta_{(5)}, {(HK^{-1})}^{\frac{1}{3}} \delta_{(10)}]\,.
\end{aligned}
\end{align}
One may verify that this is consistent with the following identifications of the internal and external metrics in the $2^6$ frame:
\begin{align}
g_{mn}^{2^6} & = \operatorname{diag}[{(HK^{-1})}^{-\frac{2}{3}}, {(HK^{-1})}^{-\frac{2}{3}}, {(HK^{-1})}^{\frac{1}{3}} \delta_{(5)}]\,,\\
{\left| g_{(4)}^{2^6}\right|}^{-\frac{1}{4}} & \coloneqq \left[ {(HK^{-1})}^{\frac{1}{6}}  {|g_{(4)}|}^{-\frac{1}{4}} \right] \Rightarrow g_{\mu \nu}^{2^6} = {(HK^{-1})}^{-\frac{1}{6}} g_{\mu \nu}\,.
\end{align}
The components of the generalised vector gets interchanged and we end up with the background of the $2^6$-brane
\begin{gather}
\begin{gathered}
\textrm{d} s^2_{2^6} = {(HK^{-1})}^{-\frac{2}{3}} ( - \textrm{d} t^2 + \textrm{d} \xi^2 + \textrm{d} \chi^2 ) + {(HK^{-1})}^{\frac{1}{3}} ( \textrm{d} z^2 + \textrm{d} {\vec{w}}^2_{(5)} ) + H^{\frac{1}{3}} K^{\frac{2}{3}} ( \textrm{d} r^2 + r^2 \textrm{d} \theta^2 )\,,\\
A_{(3)} = - H^{-1} K \textrm{d} t \wedge \textrm{d} \xi \wedge \textrm{d} \chi\,, \qquad A_{(6)} = - K^{-1} \sigma \theta \textrm{d} z \wedge \textrm{d} w^1 \wedge \ldots \wedge \textrm{d} w^5\,.
\end{gathered}
\end{gather}
One may similarly find consistent rotations of the generalised metric that give the KK6 and $0^{(1,7)}$, the details of which can be found in \cite{Berman:2018okd}. We summarise the M-theory brane configurations that we may obtain in Table \ref{tab:NonGeomM}.
\begin{table}[!ht]
\centering
\begin{tabulary}{\textwidth}{CLCCCCCCCCC}
\toprule
& & $t$ & $r$ & $\theta$ & $z$ & & $\xi$ & $\chi$ & $w^a$ &\\
\cmidrule{3-6}\cmidrule{8-10}
& $5^3$ & $\ast$ & $\bullet$ & $\bullet$ & $\circ$ & & $\circ$ & $\circ$ & $\ast$\\
& $2^6$ & $\ast$ & $\bullet$ & $\bullet$ & $\circ$ & & $\ast$ & $\ast$ & $\circ$\\
& $0^{(1,7)}$ & $\ast$ & $\bullet$ & $\bullet$ & $\circ$ & & $\odot$ & $\circ$ & $\circ$\\
& KK6 & $\ast$ & $\bullet$ & $\bullet$ & $\odot$ & & $\circ$ & $\ast$ & $\ast$\\
\bottomrule
\end{tabulary}
\caption{The configuration of M-theory branes that we consider. Asterisks $\ast$ denote worldvolume coordinates, empty circles $\circ$ denote smeared transverse coordinates and filled circles $\bullet$ denote coordinates that the harmonic function depends on. Finally, $\odot$ denotes an otherwise distinguished direction; the Hopf fibre for the monopole and the quadratic direction for the $0^{(1,7)}$}
\label{tab:NonGeomM}
\end{table}
\subsection{Reduction to IIA Solutions}
The IIA limit of EFTs is not really a distinct solution to the section constraint, being obtained from a circle compactification of M-theory, and so it is not really surprising that we may also obtain exotic branes of the Type IIA theory. Nevertheless, we briefly cover how they arise in the non-geometric solution. We take the reduction in the internal space and decompose the 7 coordinates into $y^m = (y^{\check{m}}, \eta)$ such that the extended coordinates are indexed by
\begin{align}\label{eq:IIACoords}
Y^M = (Y^{\check{m}}, Y^\eta; Y_{\check{m} \check{n}}, Y_{\check{m} \eta}; Y_{\check{m}}, Y_\eta; Y^{\check{m} \check{n}}; Y^{\check{m} \eta})\,.
\end{align}
Working in the Einstein frame\footnote{One can also work in the string frame if so preferred---this is detailed in \cite{Berman:2018okd}.}, the internal metric decomposes to
\begin{align}
g^{\text{M}}_{mn} = \operatorname{diag} [ e^{-\frac{\phi}{6}} {\check{g}}_{\check{m} \check{n}}^{\text{A}}, e^{\frac{4\phi}{3}} ]\,, \qquad g^{\text{M}}_{(7)} = e^{\frac{\phi}{3}} {\check{g}}^{\text{A}}_{(6)}\,,
\end{align}
which induces a decomposition of the generalised metric
\begin{align}\label{eq:IIAGenMetric}
\phantom{\mathcal{M}_{MN}} &
\begin{alignedat}{2}
\mathllap{\mathcal{M}_{MN}} = \smash{\underbrace{e^{\frac{\phi}{6}} {\left|{\check{g}}^{\text{A}}_{(4)} \right|}^{-\frac{1}{4}}}_{\text{from } {| g_{(4)}^{\text{M}}|}^{-\frac{1}{4}}} \operatorname{diag} [}
& e^{-\frac{\phi}{6}} {\check{g}}_{\check{m}\check{n}}^{\text{A}}, e^{\frac{4\phi}{3}}; e^{\frac{\phi}{3}} g^{\check{m} \check{n}, \check{p} \check{q}}_{\text{A}}, e^{-\frac{7\phi}{6}} g^{\check{m} \check{n}}_{\text{A}};\\
& e^{-\frac{\phi}{6}} {\check{g}}^{\text{A}}_{(6)}{}^{-1} {\check{g}}^{\check{m} \check{n}}_{\text{A}}, e^{-\frac{5\phi}{3}} {\check{g}}^{\text{A}}_{(6)}{}^{-1}; e^{-\frac{2\phi}{3}} {\check{g}}^{\text{A}}_{(6)}{}^{-1} {\check{g}}^{\text{A}}_{\check{m} \check{n}, \check{p} \check{q}}, e^{\frac{5\phi}{6}} {\check{g}}^{\text{A}}_{(6)}{}^{-1} {\check{g}}^{\text{A}}_{\check{m} \check{n}} ]\,.
\end{alignedat}
\end{align}
where we have denoted the internal and external metrics in a particular IIA frame with a superscript/subscript $\textrm{A}$ in analogy with the M-theory section before. Since the reduction is in the internal space, the external metric is unaffected and takes on the same numerical values as the M-theory case ${\check{g}}_{\check{\mu} \check{\nu}} \equiv g_{\mu \nu}$. The EFT vector decomposes to
\begin{align}
\mathcal{A}_\mu{}^M \rightarrow  (\mathcal{A}_{\check{\mu}}{}^{\check{m}}, \mathcal{A}_{\check{\mu}}{}^{\eta}; \mathcal{A}_{\check{\mu}, \check{m} \check{n}}, \mathcal{A}_{\check{\mu}, \check{m} \eta}; \mathcal{A}_{\check{\mu}, \check{m}}, \mathcal{A}_{\check{\mu}, \eta}; \mathcal{A}_{\check{\mu}}{}^{\check{m}\check{n}}, \mathcal{A}_{\check{\mu}}{}^{\check{m} \eta}).
\end{align}
As before, the $\mathcal{A}_{\check{\mu}}{}^{\check{m}}$ and $\mathcal{A}_{\check{\mu}, \check{m}}$ components sources the KK-vector in this 4+6 split and the dual graviton. Of the remaining components, the R-R potentials $C_{(1)}, C_{(3)}, C_{(5)}$ and $C_{(7)}$ are encoded in the components $\mathcal{A}_{\check{\mu}}{}^{\eta}, \mathcal{A}_{\check{\mu}, \check{m} \check{n}}, \mathcal{A}_{\check{\mu}}{}^{\check{m}\check{n}}$ and $\mathcal{A}_{\check{\mu} \eta}$ respectively (where the latter two are to be dualised on the internal space) and the NS-NS potentials $B_{(2)}$ and $B_{(6)}$ are held in $\mathcal{A}_{\check{\mu}, \eta \check{m}}$ and $\mathcal{A}_{\check{\mu}}{}^{\eta \check{m}}$.\par
The different reductions that one can be obtained from a given M-theory section is determined by the direction that one takes the reduction $\eta$. For example, if we take $\eta$ in either $\xi$ or $\chi$, one may verify that the data
\begin{align}
e^{\frac{4\phi}{3}} & = {(HK^{-1})}^{\frac{2}{3}}\,,\\
{\left|{\check{g}}^{5^2_2\text{A}}_{(4)}\right|}^{-\frac{1}{4}} & = {(HK^{-1})}^{-\frac{1}{4}} {| {\check{g}}_{(4)} |}^{-\frac{1}{4}} \Rightarrow {\check{g}}_{\check{\mu} \check{\nu}} = {(HK^{-1})}^{\frac{1}{4}} {\check{g}}_{\check{\mu}\check{\nu}}\,,\\
{\check{g}}_{\check{m} \check{n}}^{5^2_2\text{A}} & = \operatorname{diag} \left[{(HK^{-1})}^{\frac{3}{4}}, {(HK^{-1})}^{-\frac{1}{4}} \delta_{(5)} \right]\,,
\end{align}
recombined in the form \eqref{eq:IIAGenMetric}, recovers the generalised metric in the $5^3$ frame. Likewise the EFT vector decomposes as described above and we obtain the $5^2_2\text{A}$ background in the Einstein frame
\begin{gather}
\begin{gathered}
\textrm{d} s^2_{5^2_2\text{A},\text{E}} = {(HK^{-1})}^{-\frac{1}{4}} \left( -\textrm{d} t^2 + \textrm{d} {\vec{w}}^2_{(5)} \right) + {(HK^{-1})}^{\frac{3}{4}} \left( \textrm{d} z^2 + \textrm{d} \chi^2 \right) + H^{\frac{3}{4}} K^{\frac{1}{4}} \left(\textrm{d} r^2 + r^2 \textrm{d} \theta^2 \right)\,,\\
B_{(2)} = - K^{-1} \theta \sigma \textrm{d} z \wedge \textrm{d} \chi\,, \qquad B_{(6)} = - H^{-1} K \textrm{d} t^2 \wedge \textrm{d} w^1 \wedge \ldots \wedge \textrm{d} w^5\,,\\
e^{2(\phi -\phi_0)} = HK^{-1}\,.
\end{gathered}
\end{gather}
The other inequivalent reduction of the $5^3$ is obtained by choosing the reduction to be along $\eta = w^5 \equiv v$. Similarly, the other backgrounds given in Table \ref{tab:NonGeomM} reduce to the Type IIA brane configurations described in Table~\ref{tab:NonGeomIIA}.
\begin{table}[!ht]
\centering
\begin{tabulary}{\textwidth}{CLLCCCCCCCCCC}
\toprule
& & & & & & & & & & \multicolumn{2}{c}{$w^a$} & \\
\cmidrule{11-12}
& Parent & & $t$ & $r$ & $\theta$ & $z$ & & $\xi$ & $\chi$ & $u^{\textrm{a}}$ & $v$ & \\
\cmidrule{4-7}\cmidrule{9-12}
& \multirow{2}{*}{$5^3$} & $5_2^2\text{A}$ & $\ast$ & $\bullet$ & $\bullet$ & $\circ$ & & $\times$ & $\circ$ & $\ast$ & $\ast$\\
& & $4_3^3$ & $\ast$ & $\bullet$ & $\bullet$ & $\circ$ & & $\circ$ & $\circ$ & $\ast$ & $\times$\\
& \multirow{2}{*}{$2^6$} & $2_3^5$ & $\ast$ & $\bullet$ & $\bullet$ & $\circ$ & & $\ast$ & $\ast$ & $\circ$ & $\times$\\
& & $1_4^6\text{A}$ & $\ast$ & $\bullet$ & $\bullet$ & $\circ$ & & $\times$ & $\ast$ & $\circ$ & $\circ$\\
& \multirow{2}{*}{$0^{(1,7)}$} & $0_4^{(1,6)}\text{A}$ & $\ast$ & $\bullet$ & $\bullet$ & $\circ$ & & $\odot$ & $\circ$ & $\circ$ & $\times$\\
& & $0_3^7$ & $\ast$ & $\bullet$ & $\bullet$ & $\circ$ & & $\times$ & $\circ$ & $\circ$ & $\circ$\\
& \multirow{2}{*}{KK6} & $6_3^1$ & $\ast$ & $\bullet$ & $\bullet$ & $\circ$ & & $\times$ & $\ast$ & $\ast$ & $\ast$\\
& & KK5A & $\ast$ & $\bullet$ & $\bullet$ & $\odot$ & & $\circ$ & $\ast$ & $\ast$ & $\times$\\
\bottomrule
\end{tabulary}
\caption{The configuration of the Type IIA branes that we consider. A cross $\times$ denotes the direction that is being reduced on. Note that one cannot obtain the D6 as a reduction of the KK6A in this solution as the fibre lies in the external space. This is a limitation of the solution rather than of the EFT itself.}
\label{tab:NonGeomIIA}
\end{table}
\subsection{The IIB Section}
The IIB section is the only other inequivalent solution to the section condition. It is obtained by first decomposing the $\mathbf{56}$ of $E_{7(7)}$ under $\operatorname{GL}(6)\times \operatorname{SL}(2)$ to single out the 6 directions that are to combine with the 4 external directions. In this section, and in the absence of any internal potentials, the generalised metric takes the form
\begin{align}\label{eq:IIBGenMetric}
\mathcal{M}_{MN} = {|{\hat{g}}_{(4)}|}^{-\frac{1}{4}} \operatorname{diag} [ {\hat{g}}_{\hat{m} \hat{n}}; {\hat{g}}^{\hat{m}\hat{n}} \hat{\gamma}^{\hat{\alpha} \hat{\beta}}; {\hat{g}}_{(6)}^{-1} {\hat{g}}_{\hat{m} \hat{k} \hat{p}, \hat{n} \hat{k} \hat{q}}; {\hat{g}}^{-1}_{(6)} {\hat{g}}_{\hat{m} \hat{n}} {\hat{\gamma}}_{\hat{\alpha} \hat{\beta}}; {\hat{g}}^{-1}_{(6)} {\hat{g}}^{\hat{m} \hat{n}} ],
\end{align}
where $\hat{m}, \hat{n} = 1, \ldots, 6$ index the Type IIB section and $\hat{\alpha}, \hat{\beta} =1,2$ are $\operatorname{SL}(2)$ indices. Here, we also have ${\hat{g}}_{\hat{m} \hat{k} \hat{p}, \hat{n} \hat{l} \hat{q}} \coloneqq {\hat{g}}_{\hat{m} [ \hat{n}|} {\hat{g}}_{\hat{k}| \hat{l}|} {\hat{g}}_{\hat{p}|\hat{q}]}$. Finally,
\begin{align}
{\hat{\gamma}}_{\hat{\alpha} \hat{\beta}} = \frac{1}{\operatorname{Im} \tau} \begin{pmatrix} {|\tau|}^2 & \operatorname{Re} \tau\\ \operatorname{Re} \tau & 1 \end{pmatrix}.
\end{align}
is the $\operatorname{SL}(2)$ metric parametrised in terms of the axio-dilaton $\tau = C_{(0)} + i e^{-\phi}$. The base external metric remains the same as before ${\hat{g}}_{\hat{\mu}\hat{\nu}} \equiv g_{\mu \nu}$ and all the external metrics of the IIB solutions will again be proportional to it. The EFT vector splits under this subgroup as
\begin{align}
\mathcal{A}_{\mu}{}^M \rightarrow ( \mathcal{A}_{\hat{\mu}}{}^{\hat{m}}, \mathcal{A}_{\hat{\mu}, \hat{m} \hat{\alpha}}, \mathcal{A}_{\hat{\mu}}{}^{\hat{m} \hat{k} \hat{p}}, \mathcal{A}_{\hat{\mu}}{}^{\hat{m} \hat{\alpha}}, \mathcal{A}_{\hat{\mu}, \hat{m}}).
\end{align}
These will variously source KK-vectors, dual graviton components and potentials. In particular, the $\operatorname{SL}(2)$ index on the components ${\mathcal{A}}_{\hat{\mu} \hat{m} \hat{\alpha}}$ and ${\mathcal{A}}_{\hat{\mu}}{}^{\hat{m} \hat{\alpha}}$ will distinguish between the types of potentials for the 2-form and 6-forms. When $\hat{\alpha} = 1$, the potential is an R-R potential $C_{(p)}$ whilst when $\hat{\alpha} = 2$, it sources an NS-NS potential $B_{(p)}$. As one might expect, S-dual solutions are given by the interchange $1 \leftrightarrow 2$ in the solution. Finally ${\mathcal{A}}_{\hat{\mu}}{}^{\hat{m} \hat{k} \hat{p}}$, once dualised on the internal space, sources only the R-R 4-form which is a singlet under S-duality and so does not carry an $\operatorname{SL}(2)$ index.\par
We shall not go into much detail of the correspondence between the coordinates of the M-theory section and the Type IIB section but we shall mention that the 5 coordinates $w^a$ in the M-theory section enter directly into the IIB section but the final coordinate of the IIB section comes from what was one of the wrappings modes in the M-theory section (recall that taking all 6 coordinates to descend directly from the M-theory section gave the IIA section). We shall denote the Type IIB section coordinates as ${\hat{y}}^{\hat{m}} = ( \zeta, w^a)$.\par
Then it is possible to find a set of background fields $({\hat{g}}_{\hat{\mu} \hat{\nu}}^{\text{B}}, {\hat{g}}_{\hat{m} \hat{n}}^{\text{B}}, \tau)$ such that, when rearranged into the form \eqref{eq:IIBGenMetric}, gives the generalised metric according to \eqref{eq:NonGeom}. We work through all the cases in detail in \cite{Berman:2018okd}, to which we point the interested reader but, for now, summarise the brane configurations that one obtains in Table \ref{tab:NonGeomIIB}.
\begin{table}[!ht]
\centering
\begin{tabulary}{\textwidth}{LCCCCCCCCC}
\toprule
& & & & & & & & $w^a$\\
\cmidrule{8-10}
& $t$ & $r$ & $\theta$ & $z$ & & $\zeta$ & $\omega$ & ${\bar{w}}^{\bar{a}}$ & ${\ubar{w}}^{\ubar{a}}$\\
\cmidrule{2-5}\cmidrule{7-10}
$5_2^2\text{B}$ & $\ast$ & $\bullet$ & $\bullet$ & $\circ$ & & $\circ$ & $\ast$ & $\ast$ & $\ast$\\
$5_3^2$ & $\ast$ & $\bullet$ & $\bullet$ & $\circ$ & & $\circ$ & $\ast$ & $\ast$ & $\ast$\\
$3_3^4$ & $\ast$ & $\bullet$ & $\bullet$ & $\circ$ & & $\circ$ & $\ast$ & $\circ$ & $\ast$\\
$1_3^6$ & $\ast$ & $\bullet$ & $\bullet$ & $\circ$ & & $\circ$ & $\ast$ & $\circ$ & $\circ$\\
$1_4^6\text{B}$ & $\ast$ & $\bullet$ & $\bullet$ & $\circ$ & & $\circ$ & $\ast$ & $\circ$ & $\circ$\\
$0_4^{(1,6)}\text{B}$ & $\ast$ & $\bullet$ & $\bullet$ & $\circ$ & & $\circ$ & $\circ$ & $\circ$ & $\circ$\\
KK5B & $\ast$ & $\bullet$ & $\bullet$ & $\odot$ & & $\ast$ & $\circ$ & $\ast$ & $\ast$\\
\bottomrule
\end{tabulary}
\caption{The configuration of the Type IIB branes that one may construct.}
\label{tab:NonGeomIIB}
\end{table}
\section{Other Exotic Branes}\label{sec:MoreBranes}
In the previous section, we described an explicit example of an EFT accommodating exotic branes. This notion is not new in itself, being known from the DFT monopole which was shown to include the exotic `Q-brane' (smeared $5_2^2$) and `R-brane' (smeared $5_3^2$) \cite{Berman:2014jsa,Bakhmatov:2016kfn}, or $\operatorname{SL}(5)$ EFT \cite{Bakhmatov:2017les}. The number of such branes that one can describe in a single solution is EFT is obviously limited by the size of the internal space of the theory that one is working with. Moving on to larger exceptional groups, at $E_{8(8)}$, the wrapping modes of the codimension-2 exotic states that we described earlier are required to extend the space to accommodate $\rho_1 = \mathbf{248}$ of $E_{8(8)}$ EFT. Whilst we are unable to give a concrete description of what new branes one should be able to describe, it is expected that we should be able to include at least those described in \cite{Fernandez-Melgarejo:2018yxq} that were obtained by examining the multiplet structure of $E_{8(8)}$ explicitly. However, when we transition to the infinite-dimensional Kac-Moody algebras (in particular $E_{9(9)}$ and $E_{11}$), it becomes obvious that even these are not enough. We need many more---infinitely more in fact---in order to source all the wrapping directions that these theories require.\par
Additionally, a conspicuous omission in the above discussion are the D8, D9 branes in 10 dimensions, as well as the lift of the former to M-theory\footnote{This object has also been referred to as the M9 in the literature. However, we shall see that it is actually more like a KK-monopole than a M5-brane, given the mass designation $8^{(1,0)}$, which is why we have opted to call it the KK8M.} called the KK8M which have been speculated to exist for a while. These are very poorly understood codimension-1 and codimension-0 states with their signature pathological asymptotic behaviours. At codimension-2 we have the S-dual of the D7-brane, which we call the $7_3$ (also known as the NS7 in the literature), whose T-duals are rarely even mentioned in the literature.\par 
With these points in mind, it is pertinent to ask what are the full set of possible branes that one expects to find in Type II string theory or M-theory. In this section, we give a simple method to generate all the possible branes at every power of $g_s$ by recursively applying a simple algorithm to the mass formulae of the objects. Studying the mass formulae of the standard branes, it is easy to verify that duality-related objects have their masses transform according to the following rules:
\begin{align}
T_y &: R_y \mapsto \frac{l_s^2}{R_y}, \qquad g_s \mapsto \frac{l_s}{R_y} g_s\,,\label{eq:T}\\
S &: g_s \mapsto \frac{1}{g_s}, \qquad l_s \mapsto g_s^{\frac{1}{2}} l_s\,.\label{eq:S}
\end{align}
Note that T-duality preserves the power of $g_s$ of the branes, as can be seen in the D-brane chain in which all objects scale as $g_s^{-1}$ in the string frame. On the other hand, S-duality changes the $g_s$ scaling such as in the $(7_1, 7_3)$ doublet. In addition to this, we have the Type IIA--M-theory correspondence in which one object in M-theory can descend to multiple objects in Type IIA. If the object has only a simple structure, such as an M5, then the only inequivalent reductions are along the worldvolume of the brane or transverse to the brane (after an appropriate smearing to generate an isometry) and so we end up with only the $4_1$ or $5_2$ respectively. However, with a more complicated structure such as a $6^1\text{M}$, we may reduce in three ways (along the worldvolume, transverse to the brane and along the fibre) to obtain a $6_1\text{A}, 5_2^1\text{A}$ and the exotic $6_3^1\text{A}$ branes. All of these oxidations/reductions can be read off from the mass formulae simply by relating the M-theory constants to the Type IIA constants
\begin{align}\label{eq:Reduction}
\begin{rcases}
l_s & = \frac{l_p^{\frac{3}{2}}}{R_\natural^{\frac{1}{2}}}\\
g_s & = {\left( \frac{R_\natural}{l_p} \right)}^{\frac{3}{2}}
\end{rcases} \leftrightarrow
\begin{cases}
R_\natural & = l_s g_s\\
l_p & = g_s^{\frac{1}{3}} l_s
\end{cases}\,.
\end{align}
The algorithm that we propose to generate all the possible branes is the following:
\begin{itemize}
	\item Apply S-duality \eqref{eq:S} to every Type IIB object to give a new object in Type IIB
	\item Apply T-duality \eqref{eq:T} to every Type IIA/B object to give a new object in type IIB/A
	\item Lift every Type IIA object by \eqref{eq:Reduction} to give a new object in M-theory
	\item Reduce every M-theory object by \eqref{eq:Reduction} in every possible inequivalent direction to give multiple objects in IIA
\end{itemize}
Doing so generates a whole web of dualities that extends far beyond the conventional geometric sector. If one were to blindly apply these rules in the conventional string- or M-theory interpretations, then they would inevitably lead to dualising or reducing along non-isometric directions. However, within the framework of ExFT, every such transformation is permissible since the symmetry-generating transformations that we apply are blind to whether the directions being acted on are isometries or not---they simply rotate the coordinate dependence in and out of the physical spacetime.\par
The possibility of objects having a dependence on coordinates outside of the physical spacetime has been given an interpretation in string- and M-theory in the context of the GLSM in which a collection of works studying worldsheet instanton effects on the 5-brane chain \cite{Tong:2002rq,Harvey:2005ab,Kimura:2013fda,Kimura:2013zva,Kimura:2018hph} concluded that a dependence on winding modes is indicative of worldsheet instanton corrections. It is expected that this correspondence holds much more generally, and that wrapping mode dependences will analogously correspond to some instanton effects on the worldvolume. We hope that this is within reach of verifiability with current techniques.\par
This opens up the intriguing possibility of short-circuiting the rather laborious path taken in conventional discussions. Rather than having to smear a direction in the harmonic function, dualise along the isometry and then calculate the worldsheet instanton corrections to obtain the full geometry, ExFT may allow us to generate the full dual picture by allowing for winding/wrapping mode dependences from the start. Using these arguments the R-monopole, whose existence had long been speculated but whose construction had been hampered by the lack of isometries to dualise along, was constructed as a solution within the DFT monopole and the resultant background was confirmed to source an R-flux.\par
The novel contribution by ExFT is then that one should be able to construct non-trivial codimension-0 objects by rotating all coordinate dependences out of the physical section. Such states are impossible to describe in string- and M-theory as the only space-filling branes in that context must necessarily be trivial. However, the codimension-0 states that ExFT allows for have trivial structure on the spacetime that they span but non-trivial structure on the wrapping coordinates which we reinterpret as worldsheet instanton corrections upon taking section, as per the GLSM. The dependence on wrapping coordinates necessarily precludes even a locally geometric description and is thus a more general class of non-geometric objects.\par
Having argued that the algorithm given above is valid in the context of ExFT, with a possible interpretation given in string- and M-theory, we follow it to its natural conclusion. Amongst all the branes that one can generate, we find all the branes that have appeared in the literature either directly, or through the mixed symmetry potentials that they couple to. For example, we generate all of the branes that couple to the mixed-symmetry potentials that appear in \cite{Bergshoeff:2017gpw,Fernandez-Melgarejo:2018yxq,Kleinschmidt:2011vu,Lombardo:2016swq,Lombardo:2017yme}. More concretely, one may conduct an exhaustive search of all the branes that would be generated down to any power of $g_s$. We tabulate the number of distinct\footnote{We count the same brane appearing in both theories, such as the NS5A/B as different objects.} branes $N_{(\alpha)}$ that one finds at all powers down to $\alpha = -25$ in 10 dimensions, as well as all the branes in M-theory that are required to describe them, in Table \ref{tab:gsGradingOfExoticBranes}.\par
For the fourth column we have split $N_{(\alpha)}$ into the number of branes that appear only in Type IIA (denoted $A$), the number of branes that appear only in Type IIB (denoted $B$) and the number of branes that appear in both theories (denoted $C$). If $N_{(\alpha)}$ splits cleanly into $A = B = N_{(\alpha)}/2$, this means that every brane appears only in either one of the theories. The D-branes at $\alpha = -1$ follow this pattern, since Type IIA/B respectively contain only even/odd D-branes, and we consequently designate that power of $g_s$ to be of `R-R' type. Conversely, if $A=B=0$, then all the branes are common to both IIA/B and we give the designation `NS-NS'. This is seen, for example, at $\alpha = -2$ with two copies of the 5-brane chain $5_2\text{A/B} \xleftrightarrow{T} 5_2^1\text{B/A} \xleftrightarrow{T} 5_2^2 \text{A/B} \xleftrightarrow{T} 5_2^3\text{B/A} \xleftrightarrow{T} 5_2^4\text{A/B}$. In the language of DFT, the first four are the branes that couple to the generalised fluxes $F_{AB}{}^C = \{ H_{abc}, f_{ab}{}^c, Q_a{}^{bc}, R^{abc}\}$ but the final object is a less familiar codimension-0 brane called the $5_2^4$ (whose flux necessarily vanishes). This has already been proposed to exist in \cite{Kimura:2018hph} where it was presented as one of the possible solution embedded in the DFT monopole.\par
Looking at Table \ref{tab:gsGradingOfExoticBranes}, we see a clear pattern; when $n$ is odd, the set of branes is of R-R type. Additionally, when $n=2 \! \mod \, 4$, the branes are of NS-NS type. However, the situation is more complicated when $n = 0\! \mod\, 4$. These powers of $g_s$ are predominantly of NS-NS type but there is a comparatively small set of branes at those powers that break this pattern, and we have denoted this the `NS-NS violation' in the final column. It is expected that these powers of $g_s$ will house predominantly NS-NS orbits, with only a small number of R-R orbits breaking the pattern. The first instance occurring at $\alpha = -4$, contains only 3 orbits, 2 of which are NS-NS and the final one being R-R as seen in the orbits presented in \cite{Berman:2018okd}\footnote{In fact these 10 branes violating full NS-NS correspond to a T-duality chain that mirrors the D-brane chain but headed by the $9_4$ (the S-dual of the $\text{D9} = 9_1$ brane) rather than the $9_1$.}. Particularly striking is how the number of branes grows steadily as the power of $g_s$ decreases to ever more non-perturbative branes powers of $g_s$. Indeed, there is no indication that this will ever terminate and so we expect an infinite number of such exotic states which form obvious candidates for supplying the wrapping coordinates of the infinite-dimensional ExFTs.\par
\afterpage{%
\clearpage
\begin{landscape}
\begin{table}[hbtp]
\centering
\begin{tabulary}{0.7\textwidth}{cclcllclclclc}
\toprule
\multirow{2}{*}{$\alpha$} & \multirow{2}{*}{Number of Branes $N_{(\alpha)}$} & \multirow{2}{*}{Type} & \multicolumn{9}{c}{\multirow{2}{*}{Breakdown $N_{(\alpha)} = A + B + 2C$}}& NS-NS Violation \\
& & & & & & & & & & & & $A+B$\\
\midrule
0 & 4 & \llap{(}NS-NS) & $4$ & $=$ & $0$ & $+$ & $0$ & $+$ & $2$ & $\times$ & $2$ & $0$ \\
-1 & 10 & R-R & $10$ & $=$ & $5$ & $+$ & $5$ &\\
-2 & 10 & NS-NS & $10$ & $=$ & $0$ & $+$ & $0$ & $+$ & $2$ & $\times$ & $5$ &\\
-3 & 24 & R-R & $24$ & $=$ & $12$ & $+$ & $12$ &\\
-4 & 46 &  & $46$ & $=$ & $5$ & $+$ & $5$ & $+$ & $2$ & $\times$ & $18$ & $10$\\
-5 & 72 & R-R & $72$ & $=$ & $36$ & $+$ & $36$ &\\
-6 & 104 & NS-NS & $104$ & $=$ & $0$ & $+$ & $0$ & $+$ & $2$ & $\times$ & $52$ &\\
-7 & 210 & R-R & $210$ & $=$ & $105$ & $+$ & $105$ &\\
-8 & 280 & & $280$ & $=$ & $12$ & $+$ & $12$ & $+$ & $2$ & $\times$ & $128$ & $24$\\
-9 & 448 & R-R & $448$ & $=$ & $224$ & $+$ & $224$ &\\
-10 & 632 & NS-NS & $632$ & $=$ & $0$ & $+$ & $0$ & $+$ & $2$ & $\times$ & $316$ &\\
-11 & 942 &  R-R & $942$ & $=$ & $471$ & $+$ & $471$ &\\
-12 & 1244 & & $1244$ & $=$ & $36$ & $+$ & $36$ & $+$ & $2$ & $\times$ & $586$ & $72$\\
-13 & 1926 & R-R & $1926$ & $=$ & $963$ & $+$ & $963$ &\\
-14 & 2340 & NS-NS & $2340$ & $=$ & $0$ & $+$ & $0$ & $+$ & $2$ & $\times$ & $1170$ &\\
-15 & 3398 & R-R & $3398$ & $=$ & $1699$ & $+$ & $1699$ &\\
-16 & 4378 & & $4378$ & $=$ & $105$ & $+$ & $105$ & $+$ & $2$ & $\times$ & $2084$ & $210$\\
-17 & 5942 & R-R & $5942$ & $=$ & $2971$ & $+$ & $2971$ &\\
-18 & 7316 & NS-NS & $7316$ & $=$ & $0$ & $+$ & $0$ & $+$ & $2$ & $\times$ & $3658$ &\\
-19 & 10050 & R-R & $10050$ & $=$ & $5025$ & $+$ & $5025$ &\\
-20 & 12252 & & $12252$ & $=$ & $224$ & $+$ & $224$ & $+$ & $2$ & $\times$ & $5902$ & $448$\\
-21 & 16134 & R-R & $16134$ & $=$ & $8067$ & $+$ & $8067$ &\\
-22 & 19388 & NS-NS & $19388$ & $=$ & $0$ & $+$ & $0$ & $+$ & $2$ & $\times$ & $9694$ &\\
-23 & 25320 & R-R & $25320$ & $=$ & $12660$ & $+$ & $12660$ &\\
-24 & 30374 & & $30374$ & $=$& $471$ & $+$ & $471$ & $+$ & $2$ & $\times$ & $14716$ & 942\\
-25 & 38310 & R-R & $38310$ & $=$ & $19155$ & $+$ & $19155$ &\\
\midrule
M & 458124\\
\bottomrule
\end{tabulary}
\caption{Number of branes in each theory down to $g_s^{-25}$, as well as the total number of branes required in M-theory to accommodate all of them.}
\label{tab:gsGradingOfExoticBranes}
\end{table}
\clearpage
\end{landscape}
}
In \cite{Berman:2018okd}, these webs were separated out into finite T-duality orbits for the cases $\alpha \geq -7$ (the lowest power of $g_s$ that one expects to find in $E_{7(7)}$ EFT). Each orbit is characterised by a definite $g_s$ scaling and are related to orbits of differing $g_s$-scaling by S-duality and/or sharing common lifts in M-theory. For example, at $g_s^{-7}$, the 210 branes falling into 5 distinct T-duality orbits.
\section{Discussion}\label{sec:Discussion}
We have already hinted in the main text that there may be more notions of non-geometry than one might first imagine. The first signs of non-geometry are the \emph{globally non-geometric objects}, of the type that we started our discussion, such as the Q-monopole (smeared $5_2^2$). Whilst a globally geometric description eludes us as their patching generically require duality transformations, as well as the conventional diffeomorphisms and gauge transformations, one can still construct local descriptions of these objects through the supergravity fields and these are may be considered as realisations of the T-folds and U-folds proposed by Hull.\par
The second class of non-geometric objects are those that are \emph{locally non-geometric}. These are backgrounds that require a dependence on coordinates outside of the usual spacetime to describe and thus lack even a local description in terms of conventional supergravity. It is hopefully obvious that all the codimension-0 branes that we have argued must exist must necessarily be of this type. What is less obvious is that higher codimension objects can also be non-geometric in this sense---the prime example being the $5_2^3$ brane (indeed, this is the context in which this type of non-geometry was first discussed). Explicit construction of the background shows that the structure of the fields necessitates a dependence on at least one winding mode if the solution is to remain non-trivial. As discussed previously, we expect the winding mode dependence to be interpreted as worldsheet instanton corrections in the conventional supergravity lore. As such, the vast majority of the branes that we recorded are expected to be of this type.\par
To these, we add a final type of non-geometry that we shall refer to as, `truly non-geometric backgrounds'---backgrounds that are not related to any geometric background by duality transformations, thus forming entirely disconnected orbits. The very nature of how we generated our non-geometric backgrounds prevents us from probing such backgrounds and it is not obvious how one might go about constructing such backgrounds as their structure is completely unknown. Indeed, it is not clear if such examples even exist and it remains an open question if there are even more objects outside of the ones we have found.
\section*{Acknowledgements}
DSB is funded by STFC grant ST/P000754/1. ETM is supported by the Russian state grant Goszadanie 3.9904.2017/8.9 and by the Foundation for the Advancement of Theoretical Physics and Mathematics ``BASIS'' and, in part, by the program of competitive growth of Kazan Federal University. RO is supported by an STFC studentship.
\bibliographystyle{JHEP}
\bibliography{CorfuProceedings}
\end{document}